\documentclass[aps,pra,twocolumn,showpacs,superscriptaddress,floatfix]{revtex4-1}

\usepackage{graphicx}
\usepackage{color}
\usepackage{amstext}
\usepackage{amsmath}
\usepackage{booktabs}
\usepackage{array}
\usepackage{appendix}
\usepackage{bm}

\usepackage{amssymb}

\begin{document}

\title{Effective Demagnetizing Factors of \\ Diamagnetic Samples of Various Shapes}

\author{R. Prozorov}
\email[Corresponding author: ]{prozorov@ameslab.gov}
\affiliation{Ames Laboratory, Ames, IA 50011}
\affiliation{Department of Physics \& Astronomy, Iowa State University, Ames, IA 50011}

\author{V. G. Kogan}
\email{kogan@ameslab.gov}
\affiliation{Ames Laboratory, Ames, IA 50011}

\date{Submitted: 16 December 2017; accepted in Phys. Rev. Applied: 26 June 2018}


\begin{abstract}
Effective demagnetizing factors that connect the sample magnetic
moment with the applied magnetic field are calculated numerically for perfectly diamagnetic samples of various non-ellipsoidal shapes. The procedure is based on calculating total magnetic moment by integrating the magnetic induction obtained from a full three dimensional solution of the Maxwell equations using adaptive mesh. The results are relevant for superconductors (and conductors in AC fields) when the London penetration depth (or the skin depth) is much smaller than the sample size. Simple but reasonably accurate approximate formulas are given for practical shapes including rectangular cuboids, finite cylinders in axial and transverse field as well as infinite rectangular and elliptical cross-section strips.
\end{abstract}

\maketitle

\section{Introduction}

Correcting results of magnetic measurements for the distortion of the magnetic
field inside and around a finite sample of arbitrary shape is not trivial, but
necessary part of experimental studies in magnetism and superconductivity.
The internal magnetic field is uniform only in ellipsoids (see Fig.~\ref{fig2}) for which demagnetizing factors can be calculated analytically
\cite{Landau1984,Osborn1945}. In general, however, magnetic field is highly non-uniform inside
and outside of finite samples of arbitrary (non-ellipsoidal) shapes and various
approaches were used to handle the problem \cite{Joseph1965,Chen1991,Aharoni1998,Sato89,Pardo2004,Smith2010,Brandt2001}. As discussed below, the major obstacle has been that so far the total magnetic moment of arbitrary shaped samples could not be calculated and approximations and assumptions had to be made. As a result, various approximate demagnetizing factors were introduced. For example, so-called ``magnetometric" demagnetizing factor, $N_{m}$, is based on equating magnetostatic self-energy to the energy of a fully magnetized ferromagnetic prism or, more generally, considering volume-average magnetization in magnetized \cite{Chen1991,Aharoni1998,Sato89} or perfectly diamagnetic \cite{Pardo2004} media. Similarly, so-called ``fluxmetric" or ``ballistic" demagnetizing factor, $N_{f}$, is based on the average magnetization in the sample mid-plane \cite{Chen1991,Pardo2004}. In these formulations, micromagnetic calculations are used to find the distribution of surface magnetic dipoles density that satisfies the boundary conditions and the assumptions made. Then the average magnetization is calculated and used to compute the $N$ factors using formulas similar to those used in this work. One common, but generally incorrect, assumption is that the sum of demagnetizing factors along three principal axes equals to one. This is  true only for ellipsoids.
Notably, E.~H.~Brandt has used a different approach by numerically calculating the slope, $dm/dH_0$, of the magnetic moment $m$ vs. applied magnetic field $H_0$ in a perfect superconductor in the Meissner state to compute the approximate $N-$factors for a 2D situation of infinitely long strips of rectangular cross-section in perpendicular field and he extended these results to finite 3D cylinders (also of rectangular cross-section) in the axial magnetic field. We find an excellent agreement between our calculations and Brandt's results for these geometries. Also, in our earlier work, the 2D numerical solutions of the Maxwell equations obtained using finite element method were generalized to 3D cylinders and brought similar results \cite{Prozorov2000}).

Yet, despite multiple attempts, results published so far do not describe three dimensional finite samples of arbitrary non-ellipsoidal shapes to answer an important practical question: \emph{What is the \textbf{total magnetic moment} of a three dimensional sample of a particular shape in a fixed \textbf{applied} magnetic field, $\bm H_{0}$?} We answer this question by finding a way to calculate total magnetic moment from the first principles with no assumptions and introducing the \emph{effective} demagnetizing factors without referring to the details of the spatial distribution of the magnetic induction. We will first consider how these effective demagnetizing factors depend on finite magnetic permeability, $\mu_r$, which highlights the difference between ellipsoidal and non-ellipsoidal shapes. Complete treatment of finite $\mu_r $   requires separate papers in which we will focus on (a) the London-Meissner state in superconductors of arbitrary shape with finite London penetration depth and (b) demagnetizing corrections in local and linear magnetic media with arbitrary $\mu_r$.

In this work we focus on perfectly diamagnetic samples, the magnetic induction $B=0$ inside, which allows studying pure effects of sample shape. The results can be used for the interpretation of magnetic measurement of superconductors when London
penetration depth is much smaller than the sample dimensions (a good approximation almost up to $0.95T_c$) or in conducting samples subject to AC magnetic field when the skin depth is small. Our goal is to find simple to use, but accurate enough, approximate
formulas suitable for the calculations of the demagnetizing correction for many
shapes that can approximate realistic samples, such as finite cylinders and cuboids (rectangular prisms).

\section{Definitions}

In local and magnetically linear media without demagnetizing effects (infinite slab or cylinder in parallel magnetic field),
\begin{eqnarray}
  B&=&\mu   H=\mu_{0}\left( M+H\right)\,,
\label{muL}\\
 M&=&\frac{B}{\mu_{0}}-H=\chi H \label{mlinear}%
\end{eqnarray}
\noindent where $\mu_0 = 4 \pi \times 10^{-7}$ [N/A$^2$ or H/m] is magnetic permeability of free space; $\mu $ and $\chi$ are  linear magnetic permeability and
susceptibility (in general these quantities are second rank
tensors, but here we consider the isotropic case.)

It follows then that,
\begin{equation}
\chi=\frac{\mu }{\mu_{0}}-1=\mu_r-1
\label{mu}%
\end{equation}
\noindent where $\mu_r=\mu /\mu_0$ is relative magnetic permeability; $\mu_r=1$ for non-magnetic media and $\mu_r=0$ and $\chi=-1$ for a perfect diamagnet.

For finite samples of ellipsoidal shape, \emph{constant} demagnetizing factors $N$  connect the applied magnetic field $H_{0}$ along certain principle direction with the internal field, $H$,

\begin{equation}
H=H_{0}-N\,M \label{Hdemag}%
\end{equation}
\noindent and in terms of an applied field the magnetization is:
\begin{equation}
M=\frac{\chi}{1+\chi N}H_{0} \label{mdemagchiL}%
\end{equation}

In arbitrary shaped samples this simple description breaks down and we have to
introduce similarly structured effective equations, albeit \textit{applicable only for integral
quantities}. Namely, upon application of an external field $H_{0}$, a finite
sample of a given shape develops a measurable total magnetic moment $m$. We now \textit{define} ``effective" (or ``integral", or ``apparent") magnetic susceptibility, $\chi_0$, and corresponding demagnetizing factor, $N$, by writing relations structurally similar to Eq.~(\ref{mdemagchiL})
\begin{equation}
m=\chi_0 H_{0} V = \frac{\chi H_0 V}{1+\chi N} \label{mom}%
\end{equation}
\noindent which reduces to conventional equations in the case of a linear magnetic material of ellipsoidal shape. Importantly, Eq.~(\ref{mom}) contains only one property to be determined - intrinsic susceptibility, $\chi$ provided the demagnetizing factor $N$ can be calculated for given geometry. This can be done for model materials with known (assumed) $\chi$ and numerically evaluated $\chi_0$ by inverting Eq.~(\ref{mom}) to obtain:

\begin{equation}
N=\frac{1}{\chi_0}-\frac{1}{\chi} \,. \label{N}%
\end{equation}
\noindent where $-1 \leq \chi \leq\infty$ and $-\infty\leq\chi_0\leq\chi$. To eliminate the influence of the material, for calculations of $N$ we will consider a perfect diamagnet with $\chi=-1$, so that when $\chi_0=\chi=-1$, $N=0$ as it should be in case of no demagnetizing effects (infinite slab or a cylinder in parallel field) and $N\rightarrow1$ for infinite plate in perpendicular field where $\chi_0\rightarrow -\infty$, while $\chi$ is still equals -1.

The main issue in using Eq.~(\ref{N}) to calculate demagnetizing factors is to calculate the total magnetic moment of a sample of a given (arbitrary) shape. There are two ways of approaching this. In non-magnetic (super)conductors, one first solves Maxwell equations with the help of one of existing numerical software packages (such as COMSOL, \cite{COMSOL}), to find the transport current density $\bm{j}(\bm{r})$. Then, the total magnetic moment is given by \cite{Landau1984},
\begin{equation}
\bm{m}=\frac{1}{2}\int\left[  \bm{r}\times\bm{j(\bm{r})}\right]  dV \label{MdefJ}%
\end{equation}
\noindent The integral here can be evaluated over the entire space, but the integrand is non-zero only inside the sample where the currents flow.

The second way to calculate the total magnetic moment, $\bm{m}$, is given by,
\begin{equation}
\bm{m}=\alpha\int \left[  \frac{\bm{B}(\bm{r})}{\mu_{0}}-\bm{H}_{0}\right]  d^3\bm{r}, \label{MdefB}%
\end{equation}
\noindent where $\bm{B}(\bm{r})$ is the actual field and $\bm{H}_0$ is the uniform applied field. Here the integral must be evaluated in a region  that includes the sample (can be the entire space). We show in the next section that this integral is not unique, but depends on the way chosen for the integration. This is accounted for by a constant $\alpha$ in Eq.~(\ref{MdefB}), $\alpha=3/2$ for integration over the large spherical domain that includes the whole sample, whereas $\alpha=1$ for integration domain as a large cylinder with the axis parallel to $\bm H_0$. It turns out that for numerical reasons, the cylindrical domain is preferable and we used it for our numerical work. Equation (\ref{MdefB}) is central to the present work, because it allows calculations without using the current distribution.
This equation (with $\alpha=3/2$) can  be found in Jackson's textbook,   Ref.~[\onlinecite{Jackson2007}], Eq.~(5.62). A related discussion about the multipole representation of the field outside the region where the field sources are localized is given in Ref.~[\onlinecite{Morse}].

For evaluation of $\bm B(\bm r)$, one can use approximation of a fully diamagnetic sample imposing ``magnetic shielding" boundary conditions available in the COMSOL software. Employing Eq.~(\ref{MdefB}) with $\bm{B}(\bm{r})$ simplifies numerical procedure and improves accuracy considerably. However, proving Eq.~(\ref{MdefB}) is not at all trivial and we derive it analytically in the next section. We also verified the results by calculating total magnetic moment $\bm{m}$ utilizing both approaches evaluating current distribution in superconducting samples using London equations and employing Eq.~(\ref{MdefJ}), and using COMSOL generated field distribution $\bm{B}(\bm{r})$ and Eq.~(\ref{MdefB}).

\section{Total magnetic moment $\bm m$}

According to Jackson's book \cite{Jackson2007}, the magnetic moment $\bm m$ of current distribution induced by an applied uniform field $\bm H_0$   in a finite sample, is related to the distribution of the magnetic induction $\bm{B}(\bm{r})$ by
\begin{eqnarray}
 { \bm I }=\int\limits_{\cal R}\left[\frac{{\bm B}(\bm r)}{\mu_0}-{\bm H}_0\right]d^3\bm r =\frac{2}{3}\,\bm{m}\,.
  \label{eq0}
\end{eqnarray}
\noindent where $\cal R$ is a radius of a large sphere containing the whole sample. In particular, ${\cal R}$ can be infinite, e.g. the integral can be extended to the whole space. This relation is central for our calculations, so that we provide a more general derivation than  that given in \cite{Jackson2007}. We show that depending on the way chosen to evaluate the integral $\bm I$ over the whole space, Eq.~(\ref{MdefB}) can have different forms, parameterized by a factor $\alpha$.

The field $\bm B$ consists of the applied field  $\bm{H}_0$  and the field $\bm h$ due to currents $\bm j$ in the sample of a finite volume $V$:
\begin{eqnarray}
\frac{{\bm B}}{\mu_0}={\bm H}_0 +\bm h\,,
  \label{eq1}
\end{eqnarray}
\noindent i.e.,  $\bm I=\int {\bm h}\, d^3\bm r$, where according to Biot-Savart law,
\begin{eqnarray}
  {\bm h}(\bm r)=\frac{1}{4\pi}\int\limits_V d^3{\bm \rho}\,\frac{{\bm j}(\bm \rho)\times{\bm R}}{R^3}\,,\qquad \bm R=\bm r -\bm \rho\,.
  \label{eq2}
\end{eqnarray}
Hence, we have
\begin{eqnarray}
  4\pi\bm I&=& \int\limits_{\cal R} d^3 {\bm r} \int\limits_V d^3{\bm \rho}\frac{{\bm j}(\bm \rho)\times{\bm R}}{R^3}\nonumber\\
 &=&
\int\limits_V d^3{\bm \rho} \,{\bm j}(\bm \rho)\times \int\limits_{\cal R} d^3\bm r \frac{{\bm R}}{R^3}\nonumber \\
 &=&
 \int\limits_V d^3{\bm \rho} \,{\bm j}(\bm \rho)\times{\cal E}(\bm \rho)\,.
  \label{eq3a}
\end{eqnarray}
\noindent where we introduce ``pseudo-electric field", ${\cal E}(\bm \rho)=  \int  d^3\bm r  {\bm R}/R^3$, which is analogous to the electrostatic field of a uniform charge distribution with a constant density of $-1$ in the whole space.
For $\bm \rho=0$, we must have  ${\cal E}=\int d^3\bm r  ({\bm r}/r^3)=0$  by symmetry.
For such a distribution, the field  ${\cal E}$ is not defined uniquely, it \emph{depends on the way one divides the space in charged elements}.

If one uses elements as spherical shells, and applies the Gauss theorem to a sphere of a radius $\bm\rho$ one obtains:
 \begin{eqnarray}
{\cal E}=-\frac{4\pi}{3}\,\bm\rho \,.
  \label{eq8}
\end{eqnarray}
Hence, we have
 \begin{eqnarray}
  \bm I= -\frac{ 1}{3 }\int\limits_V d^3{\bm\rho} \,{\bm j}(\bm \rho)\times {\bm \rho}=\frac{2}{3}\, {\bm m} \,,
  \label{eq9}
\end{eqnarray}
\noindent where $\bm m$ is the total  magnetic moment. It is worth noting that this formula holds for any current distribution within the finite sample of arbitrary shape.

If one uses integration elements as cylindrical shells parallel to $\bm H_0$, i.e. choose the volume element as $2\pi \rho_1\,d\rho_1dz$ ($\bm \rho_1$ is the cylindrical radius vector), and applies the Gauss theorem to a cylinder of a radius $\rho$ one obtains:

\begin{eqnarray}
{\cal E}  =- 2\pi {\bm \rho}_1 \,.
  \label{eq8a}
\end{eqnarray}
Substituting this in Eq.~(\ref{eq3a}), one expresses the $z$ component of the integral $\bm I$:
 \begin{eqnarray}
   I_z=    {m}_z \,.
  \label{eq9a}
\end{eqnarray}

It is easy to show that the region where the integral {\bf I} is evaluated can be taken as a sphere (or a cylinder) of a radius ${\cal R}_1$ that contains the entire sample of interest within this region. Then, if one takes a larger radius ${\cal R}_2$, the layer between  spheres (cylinders) ${\cal R}_1$ and ${\cal R}_2$ does not contribute to the effective field ${\cal E}$ because the ``electric field" of such a uniformly charged spherical (cylindrical) shell $ {\cal E} (r)=0$ for $r<{\cal R}_1$.

In  Appendix A, for demonstration purposes, we evaluate the integral $\bm{I}$ both analytically and numerically for spherical and cylindrical integration volumes for the case of a current ring   for which the distribution of $\bm{B}(\bm{r})$ is known.

 It is worth mentioning that a similar argument can be applied for evaluation of the dipole moment of a metallic  sample of arbitrary shape placed in a uniform electric field $\bm E_0$, $\bm d =\int d^3\bm \rho \,n(\bm \rho) \bm \rho$, ($n(\bm \rho)$ is the charge density, for point charges $\bm d =\sum_\nu   e_\nu  \bm \rho_\nu$). It is straightforward to see that
 \begin{eqnarray}
   \bm d\propto \int d^3\bm \rho \,\left[\bm E (\bm \rho )- \bm E_0\right] \,.
  \label{eq19a}
\end{eqnarray}

Here, $\bm E (\bm \rho )$ is the electric field distribution which can be found numerically with the help of a software similar to COMSOL \cite{COMSOL}. As in the magnetic case, the integration here can be done over a spherical (or cylindrical) region which contains the whole sample. The coefficient of proportionality here is $2 \alpha/ \epsilon_0$, where $\alpha$ is given above and $\epsilon_0$ is the vacuum dielectric constant. This result might be useful in problems like those considered in \cite{polarization}.

In Appendix B we provide a derivation of Eq.\,(\ref{MdefB}) in Gaussian units for readers who prefer CGS in general electromagnetic problems.

\section{Numerical calculations}

Numerical calculations of three dimensional distribution of vector   $\bm{B}(\bm{r})$, were performed with COMSOL software \cite{COMSOL} using adaptive finite element solution of the Maxwell equations in the form applicable to many different situations, including external currents $\bm{j}_{ext}$ and electrical conductivity $\sigma$ in case of conducting materials.

\begin{equation}
\begin{aligned}
&\nabla \times \bm{H}=\bm{j},\\
&\bm{B}=\nabla \times \bm{A},\\
&\bm{j}=\sigma \bm{E} + \bm{j}_{ext},\\
&\bm{B}=\mu_0 \mu_r \bm{H},
\end{aligned}
\label{ampere}
\end{equation}
We used finite $\mu_r$ below to illustrate $\mu_r$-dependent effective demagnetizing factor. Otherwise, throughout the manuscript, $\mu_r=0$, $\sigma=0$ and $\bm{j}_{ext}=0$.
For details the reader is referred to extensive documentation available on COMSOL web site \cite{COMSOL}. ``AC/DC magnetic field" COMSOL module in a stationary DC study was used to model perfect diamagnetic material. Frequency-dependent AC study was used to formulate London equations with complex frequency-dependent conductivity. In the limit of perfect diamagnetic material both approaches gave identical results.

The main numerical difficulty is to construct the proper adaptive mesh, which should be fine enough to resolve surface currents, but still give solutions in reasonable time. Various strategies were emploied to optimize the process, utilizing symmetries, periodic boundary conditions, perfect magnetic shielding, and various adaptive sweeps and batch modes. Each geometry was solved for by using several different approaches and different meshes to make sure final results are model-independent. Geometries for which analytical solutions are known (ellipsoids and cylinders) were used to verify numerical schemes and gave nearly perfect agreement. All calculations were done in SI, so that factor $\mu_{0}$ was properly taken into account where required.

\begin{figure}[tbh]
\includegraphics[width=1\linewidth]{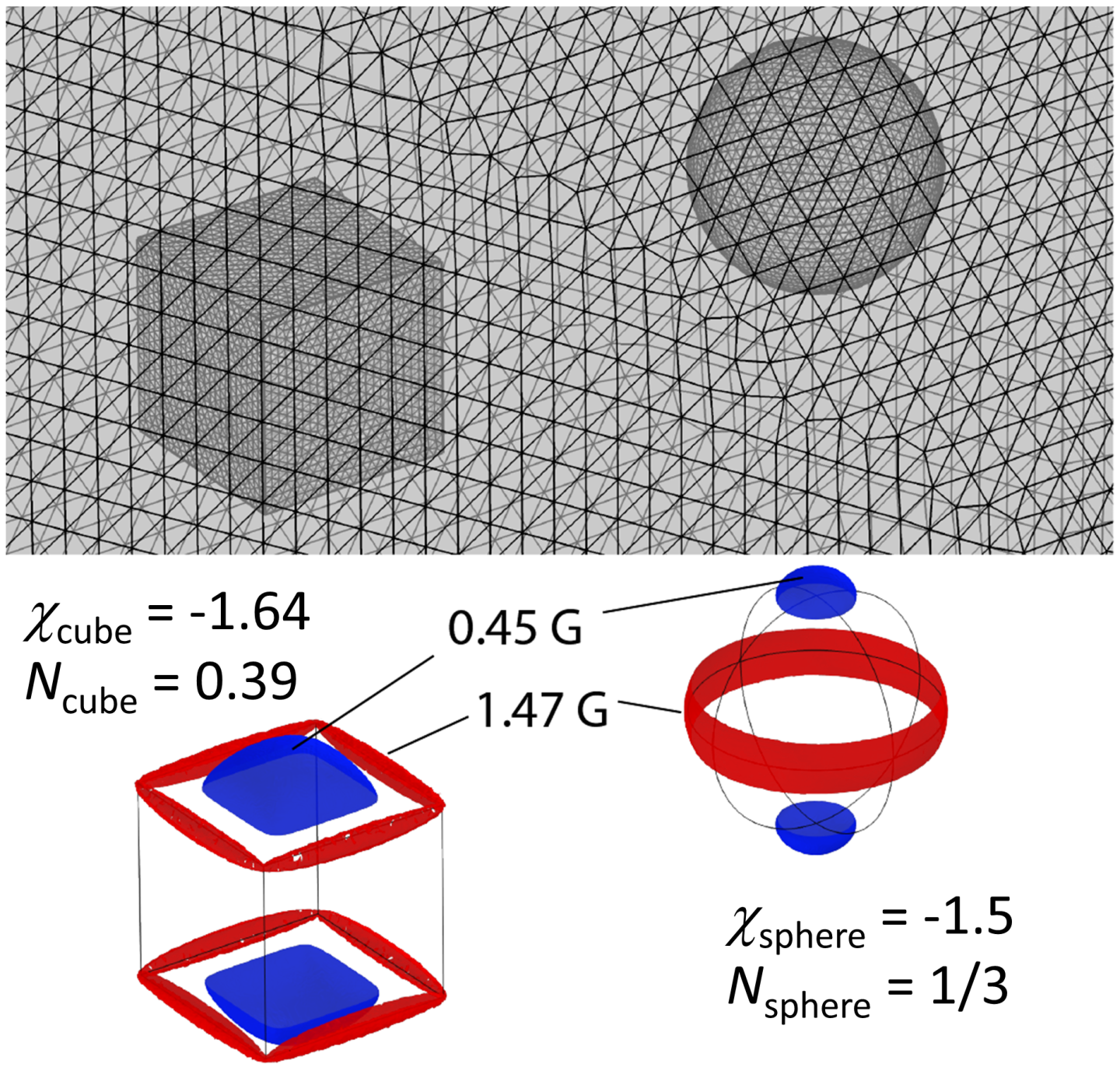}%
\caption{(top panel) sphere and cube in a full three dimensional meshed model.
(bottom panel) corresponding 3D solutions showing surfaces of constant amplitude
magnetic induction of 0.45 G and 1.47 G. Applied external field is  1
Oe.}
\label{fig1}
\end{figure}

To illustrate the method, top panel of Fig.\ref{fig1} shows three-dimensional
meshed sphere (right) and cube (left). The meshes used in actual calculations
were much finer and contained various adaptive refinements and layers. (they
would be irresolvably dark if shown here). Bottom panel of Fig.\ref{fig1}
shows two surfaces of constant magnetic induction around these samples. With
the applied field of $1$ Oe $\approx 79.58$ A/m, one surface with $0.45$ G $= 0.045$ mT corresponds to
diamagnetic shielding outside the sample, while $1.47$ G $= 0.147$ mT corresponds to
enhancement due to demagnetization. Clearly, cube provides more shielding, $\chi_0=\chi_{cube}=-1.64$,
compared to the sphere, $\chi_0=\chi_{sphere}=-1.5$, and this is reflected in a larger demagnetizing
factor, $N_{cube}=0.39$ compared to $N_{sphere}=1/3$. Already here, it is
obvious that, due to symmetry, the sum of demagnetizing factors in three
principal directions for a cube is $\sum N_{i}=3\times0.39=1.17>1$.

\section{Finite magnetic permeability}

Unfortunately, complications arise in non-ellipsoidal samples with finite
magnetic permeability. While demagnetizing factors are constants
independent of $\mu_r$ in ellipsoidal samples, they become $\mu_r$-dependent
otherwise. Hence, effective demagnetizing factors are no longer purely geometric parameters. It is still possible to provide some practical approximation of this behavior, but it will require a separate paper.%

\begin{figure}[tb]
\includegraphics[width=1\linewidth]{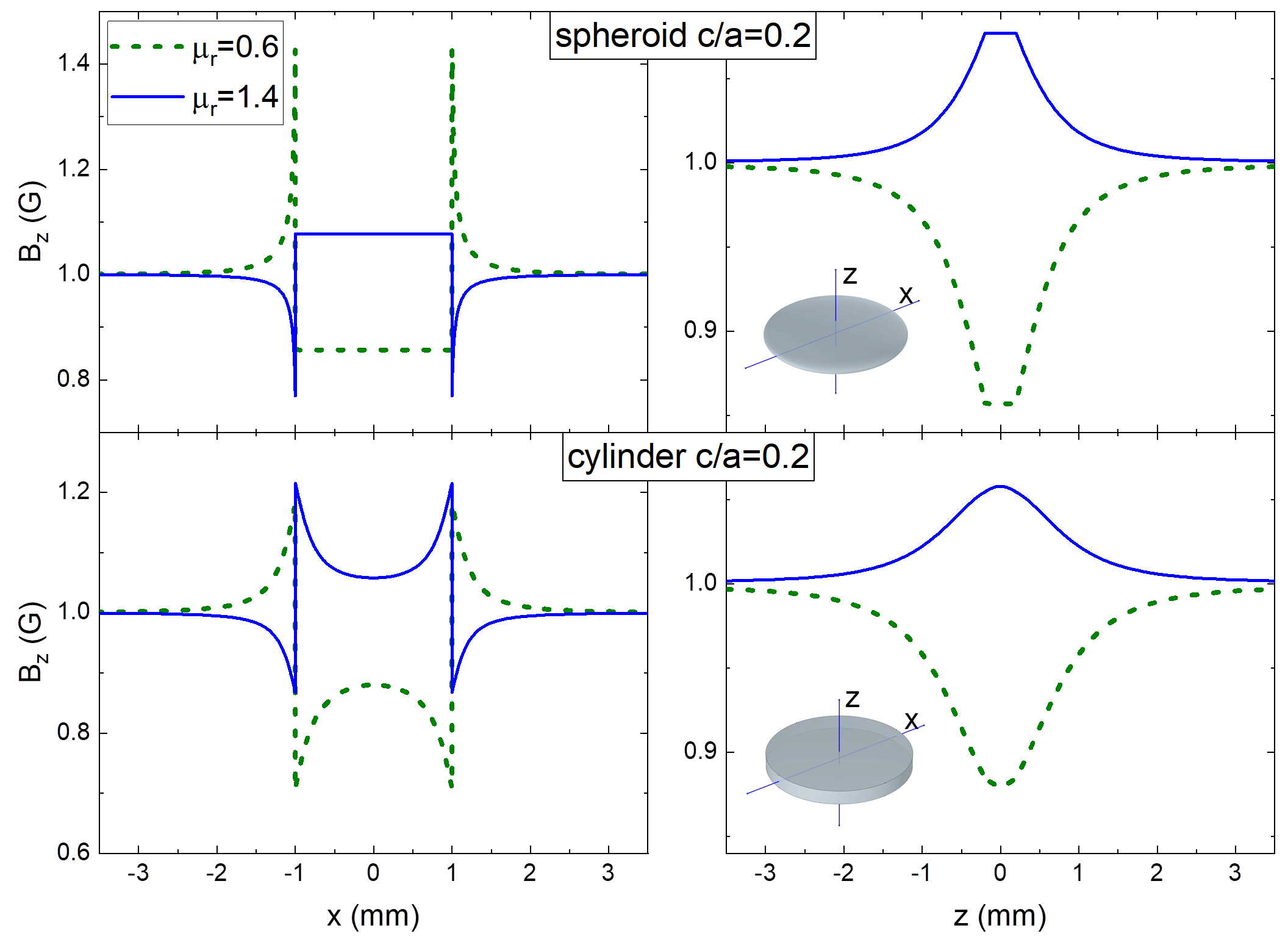}%
\caption{$B_{z}$ component of the magnetic induction across the sample in the $x-$
direction (left panels) and in the $z-$ direction (right panels) for two values of
relative magnetic permeability, $\mu_r=0.6$ (diamagnetic, dashed lines) and
$\mu_r=1.4$ (paramagnetic, solid lines). Top panels are for an oblate spheroid
and bottom panels are for a cylinder of the same aspect ratio (see insets).
Note constant field inside a spheroid and strongly non-uniform magnetic induction inside a cylinder.}
\label{fig2}
\end{figure}

Here we outline all the steps of calculating effective demagnetizing factors. First
we use COMSOL to calculate three dimensional distribution of the vector $\bm{B}(\bm{r})$ inside and outside the sample. Here we do it for two values of magnetic permeability corresponding to diamagnetic and paramagnetic materials.
Figure \ref{fig2} shows $B_{z}$ component of the magnetic induction across the
sample in the  $x$ direction, $B_{z}\left(  x,z=0\right)$, (left panels) and in the
$z$ direction $B_{z}\left( x=0,z\right)  $ (right panels) for two values of
relative magnetic permeability, $\mu_r=0.6$ (diamagnetic, dashed lines) and
$\mu_r=1.4$ (paramagnetic, solid lines). Top panels are for an oblate spheroid
and bottom panels are for a cylinder of the same aspect ratio (see insets).
Note constant magnetic induction inside a spheroid and a very non-uniform induction inside a cylinder.

The next step is to use Eq.(\ref{MdefB}) with spherical or cylindrical integration volumes to compute total magnetic moment in a fixed applied field of 1 Oe. Finally, we use Eq.(\ref{N}) to evaluate the effective demagnetizing factor.

\begin{figure}[tb]
\includegraphics[width=1\linewidth]{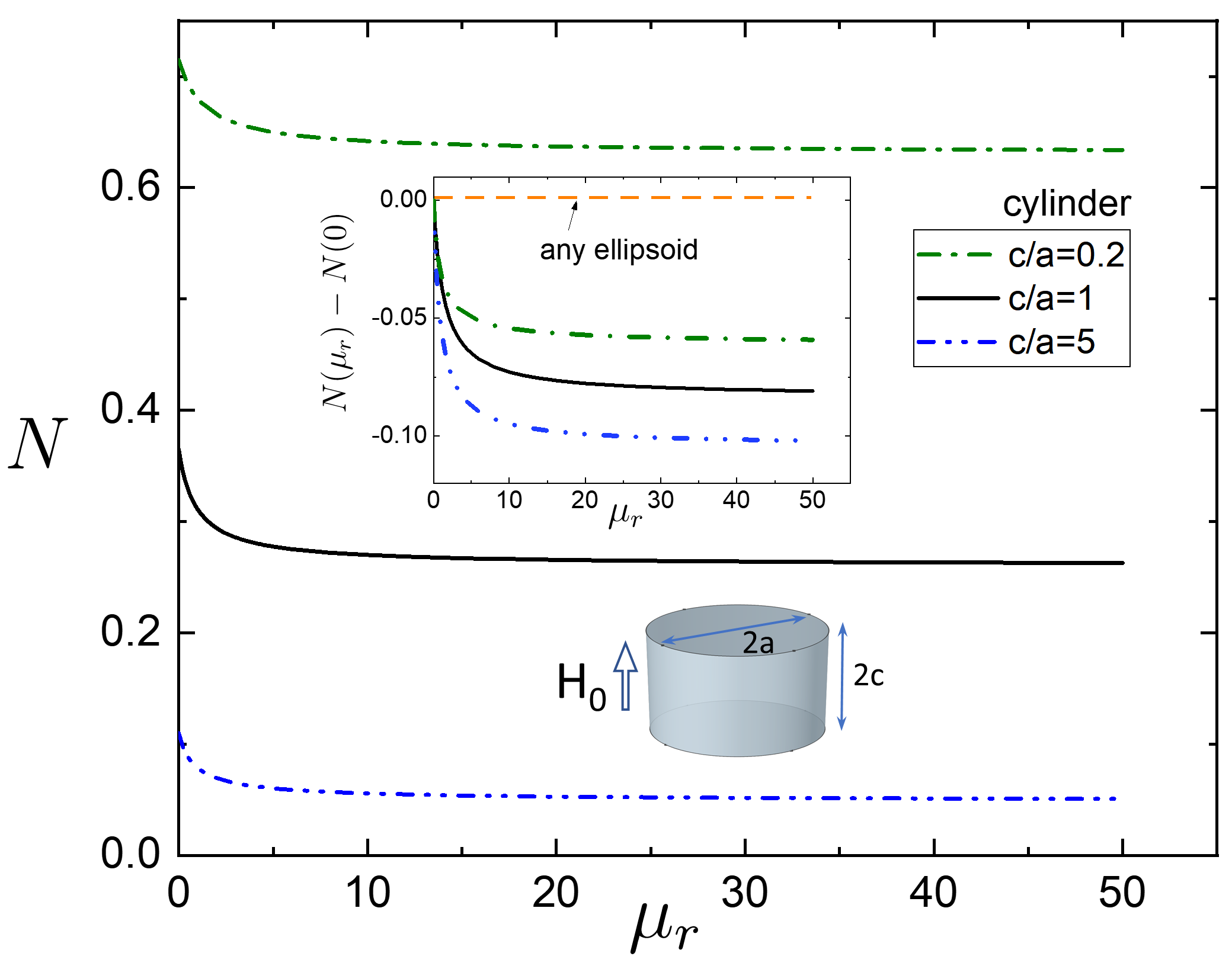}%
\caption{Effective demagnetizing factor $N$ for a finite cylinder in an
axial magnetic field for three different aspect ratios as a function of relative magnetic permeability $\mu_r$. Inset shows the
difference $N\left(  \mu_r\right) - N\left(  0\right)$.}
\label{fig3}
\end{figure}

Figure \ref{fig3} shows $\mu_r$ dependence of the effective demagnetizing
factor $N (\mu_r)$ of a finite cylinder in longitudinal magnetic field for three different values of the thickness to radius aspect ratio. The
inset shows the variation of the difference $N\left(  \mu_r\right)  -N\left(
0\right)$. Expectedly, (see Eq.(\ref{N})), for a strongly  paramagnetic material with $\mu_r>5-10$ the
variation in $N$ is not too substantial. However, for materials of practical interest,
$0\leq\mu_r \leq10$ the dependence of $N$ on $\mu_r$ is quite strong. We will attempt to
provide a simplified description of $N\left(  \mu_r \right)  $ for various
non-ellipsoidal shapes elsewhere.

\section{Perfect Diamagnets}
For now we will focus on a perfect diamagnetic material with constant $B=0$ inside.

\subsection{General ellipsoid}

Throughout this paper we adapt uniform designation of sample dimensions
$2a\times2b\times2c$ along Cartesian $x,y$ and $z$ axes with external magnetic
field applied along the $z-$ axis, parallel to the side $c$ of the sample.
Also, we will always use the dimensionless ratios, $b/a$ and $c/a$.

For completeness, it is useful to show here the analytical solution for the ellipsoid with semi-axes,
$a$, $b$ and $c$ given in Ref.~\onlinecite{Landau1984}, Eqs.(4.5) and (4.25). Osborn also gives analytical solutions of this case
expressed via the differences of incomplete elliptic integrals and written for
a restricting case of $a\geq b\geq c$ \cite{Osborn1945} (Eqs.(2.1-2.6)). It turns out formulas given in Landau's
book are much easier to compute numerically and they work for any ratio of
the dimensions \cite{Landau1984}. The demagnetizing factor along the $c$ axis is:

\begin{equation}
N_{ellipsoid}=\frac{1}{2}\frac{b}{a}\frac{c}{a}\displaystyle\int\limits_{0}^{\infty}
\frac{ds}{\left(  s+\frac{c^2}{a^2}\right)  R\left(  s\right)  }
\label{Nellipsoid}
\end{equation}

\noindent where,
\begin{equation}
R\left( s\right)  =\sqrt{\left(  s+1\right)  \left(  s+\frac{b^2}{a^2}\right)  \left(  s+\frac{c^2}{a^2}\right)}
\label{Rs}%
\end{equation}

Demagnetizing factors along other two directions have similar form with
$\left(  s+(c/a)^2\right)  $ in the denominator of Eq.(\ref{Nellipsoid}) replaced by
$\left(  s+1\right)$ or $\left(  s+(b/a)^2\right)$, along $a$ axis or $b$ axis, respectively. We verified our numeric
approach by calculating $N$ for ellipsoids and found a perfect agreement with Eq.(\ref{Nellipsoid}).

\subsection{Rectangular cuboid}

Brick-shaped sample most commonly
encountered in research laboratories, because many single crystals tend to
grow in this shape, cutting and polishing procedure also favors this type
of samples.%

\begin{figure}[tb]
\includegraphics[width=1\linewidth]{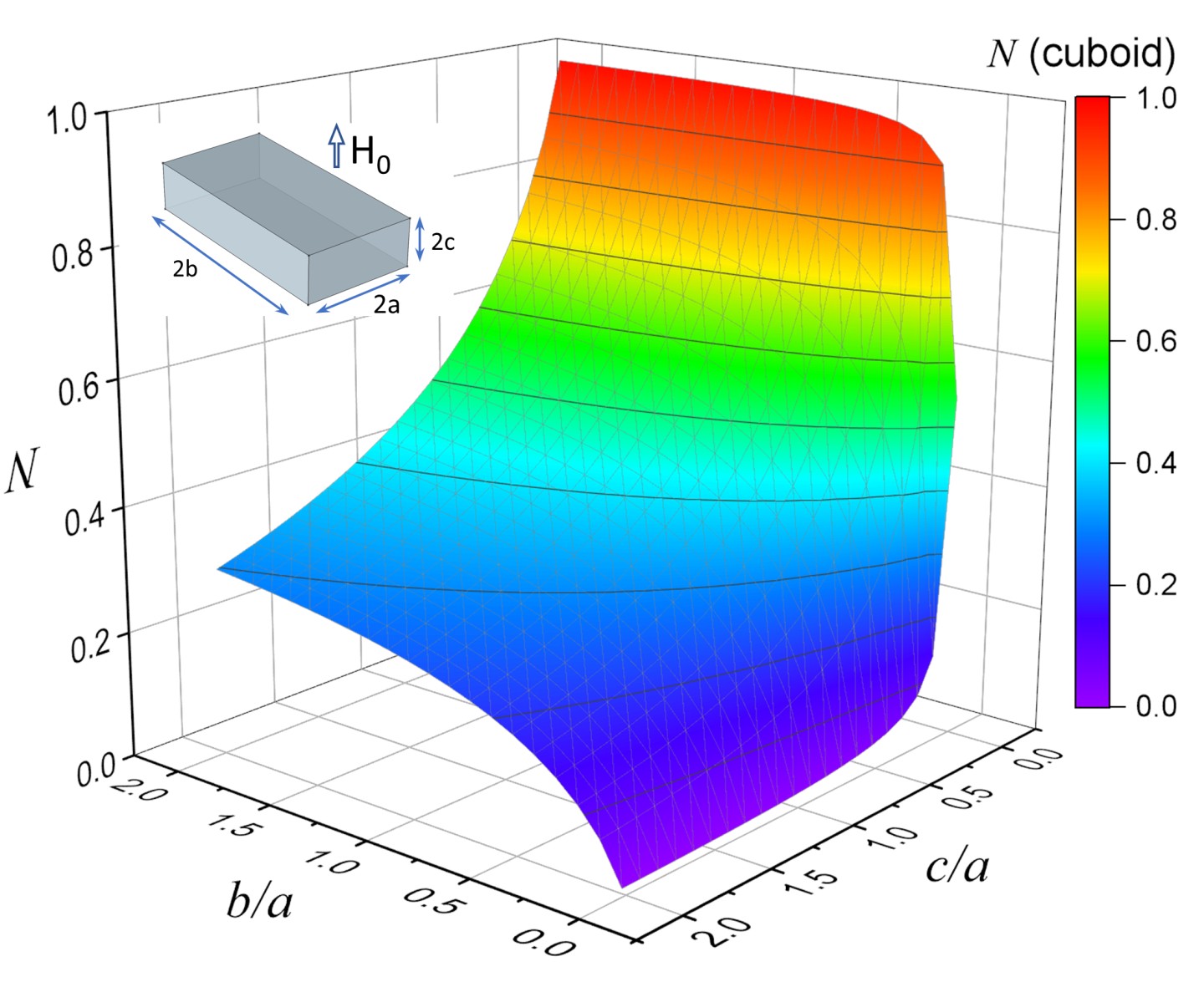}%
\caption{Effective demagnetizing factor of a rectangular cuboid as function
of its two aspect ratios. Inset shows schematic geometric arrangement.}
\label{fig4}
\end{figure}

Three-dimensional surface of the effective demagnetizing factor, $N$, of a
rectangular cuboid as function of two aspect ratios, $c/a$ and $b/a$, is shown
in Fig.\ref{fig4}. Analysis of the numerical data led us to suggest the simple
formula,
\begin{equation}
N_{cuboid}\approx\frac{4ab}{4ab+3c\left(  a+b\right)  }
\label{Ncuboid}%
\end{equation}
\textit{This is an important result of this work}, because it describes most frequently used sample shape.
Note that Eq.(\ref{Ncuboid}) is an interpolation between limiting cases of
infinitely thin sample ( $c\rightarrow0,$ $N\rightarrow1$) and $N\rightarrow0$
in the opposite case.%

\begin{figure}[tb]
\includegraphics[width=1\linewidth]{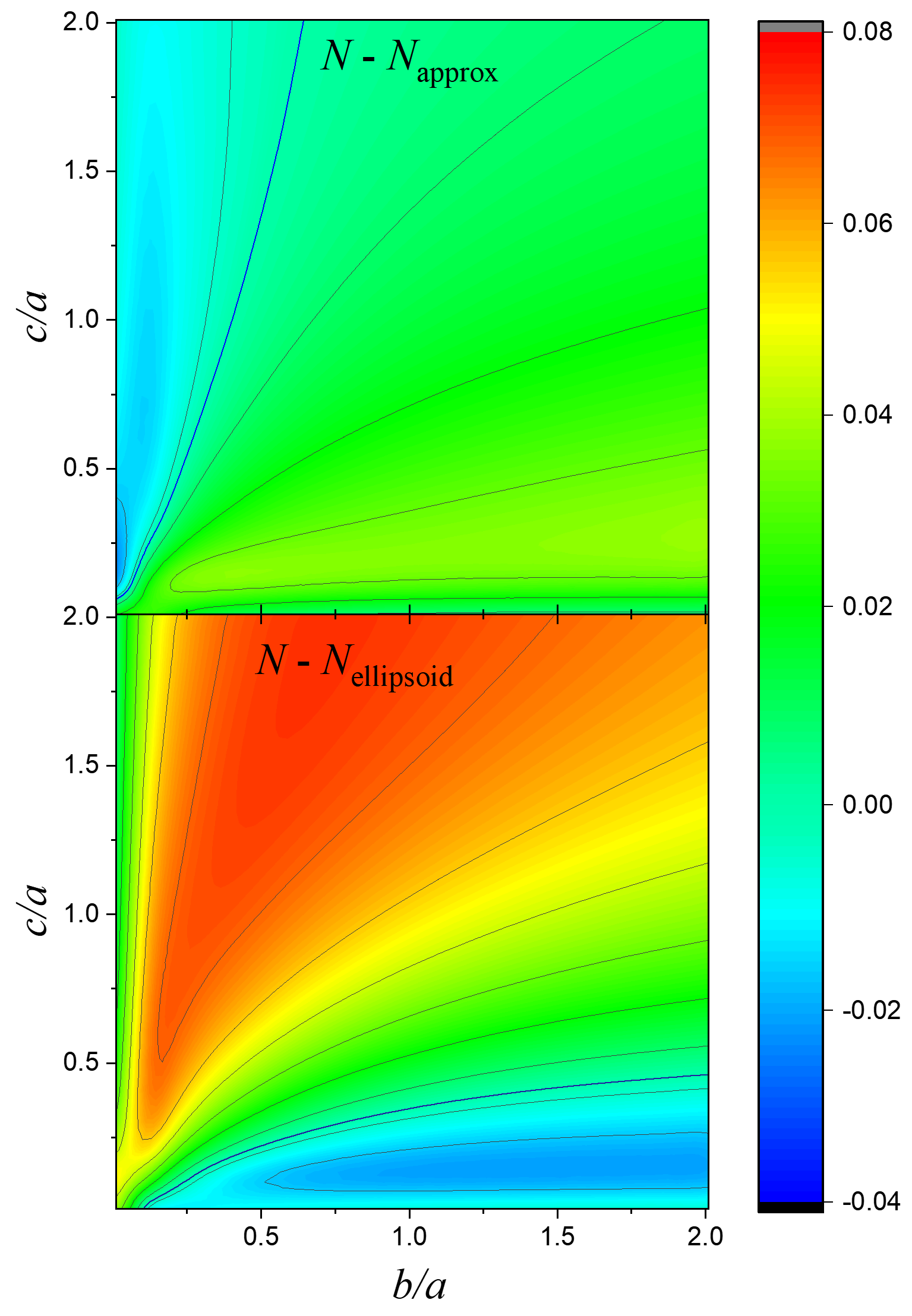}%
\caption{Top panel: the difference between numerically calculated $N$ and
approximation given by Eq.(\ref{Ncuboid}). Bottom panel: the difference
between numerically computed $N$ for cuboid and that of an ellipsoid,
Eq.(\ref{Nellipsoid}).}
\label{fig5}
\end{figure}

Top panel of Fig.~\ref{fig5} shows the difference between numerically
calculated $N$ for a rectangular cuboid (Fig.~\ref{fig4}) and approximation
given by Eq.(\ref{Ncuboid}). For comparison, bottom panel of
Fig.\ref{fig5} shows the difference between numerically computed $N$ for
cuboid and analytical solution for an ellipsoid, Eq.(\ref{Nellipsoid}). While the
latter shows deviation upward of $0.08$ (considering that $N$ can only vary
between $0$ and $1$), the former remains a much closer function approximating numerical results.

The relative error in determining magnetic moment (therefore apparent susceptibility) of a finite sample due to non-exact demagnetizing factor is readily derived from Eqs.~\ref{mdemagchiL}) and (\ref{N}),
\begin{equation}
\frac{m(N)-m(N_{approx})}{m(N)}=\frac{N_{approx}-N}{N_{approx}+\chi^{-1}}
\label{chiErr}%
\end{equation}
\noindent where $N$ is the exact and $N_{approx}$ is the approximate demagnetizing factors, respectively and $\chi$ is intrinsic magnetic susceptibility. Estimates using Eq.(\ref{chiErr}) show that most of the diagram (except for very thin samples) in Fig.~\ref{fig5}(a) results in errors within 10\%. The error becomes larger for $b/a \gtrsim 1$ and $c/a \lesssim 0.1$ (lower right) corner of Fig.~\ref{fig5}(a). Indeed, if better precision is needed, full calculations are required.

\begin{figure}[tb]
\includegraphics[width=1\linewidth]{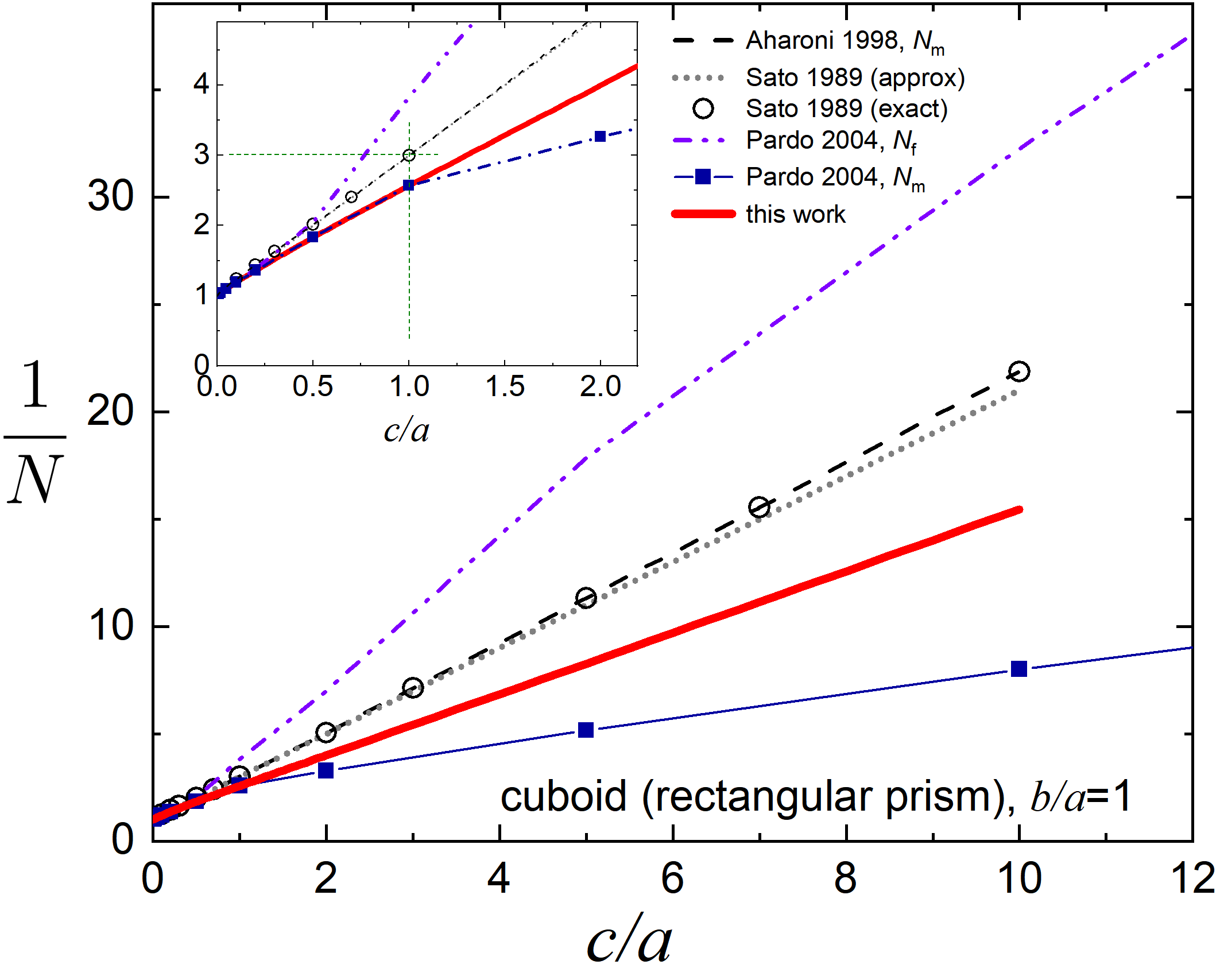}%
\caption{Comparison of the effective demagnetizing factors for a square base
cuboid: this work (solid red line), ``magnetometric", $N_m$, from Aharoni \cite{Aharoni1998} (dashed black line), ``magnetometric", $N_m$ (solid squares) and ``fluxmetric", $N_f$, (dash double dotted line) from Pardo {\it et al.} \cite{Pardo2004}, and magnetometric $N$ - factors, - approximate (dotted grey), exact (circles) from Sato and Ishii \cite{Sato89}. Inset zooms at thinner samples. Of all calculations, magnetometric $N$ factors calculated by Pardo {\it et al.} \cite{Pardo2004} are in a very good agreement with out results for $c \leq a$.}
\label{fig6}
\end{figure}

By matching magnetostatic self-energy to total magnetic energy of a saturated ferromagnetic prism, Aharoni has provided formulas for so-called ``magnetometric" demagnetizing factor, $N_m$ of the rectangular prism \cite{Aharoni1998}. Similarly, Pardo {\it et al.} calculated both ``magnetometric" factor using volume average magnetic field and ``fluxmetric" factors using mid-plane average magnetic fields \cite{Pardo2004}. Both postulated that the sum of three demagnetizing factor must be $1$ which is not justified. However, when Pardo {\it et al.} relaxed this constraint calculating their ``magnetometric" $N$-factors, their results agreed perfectly with our numerics for $c \leq a$ \cite{Pardo2004}. Sato and Ishii provided very simple approximate formulas for square cuboids and circular cylinders of finite thickness in axial magnetic field  \cite{Sato89} that they obtained by the analysis of the solutions by Aharoni and co-workers. They, therefore, agree well with Aharoni  \cite{Aharoni1998}, but disagree with our unconstrained numerical results. This is shown graphically in Fig.~\ref{fig6} where various effective demagnetizing $N$ - factors are shown for a square base cuboid as function of thickness to side ratio, $c/a$.

\subsection{Finite cylinder in axial and transverse magnetic fields}
\subsubsection{Finite circular cylinder in an axial magnetic field}
\begin{figure}[tb]
\includegraphics[width=1\linewidth]{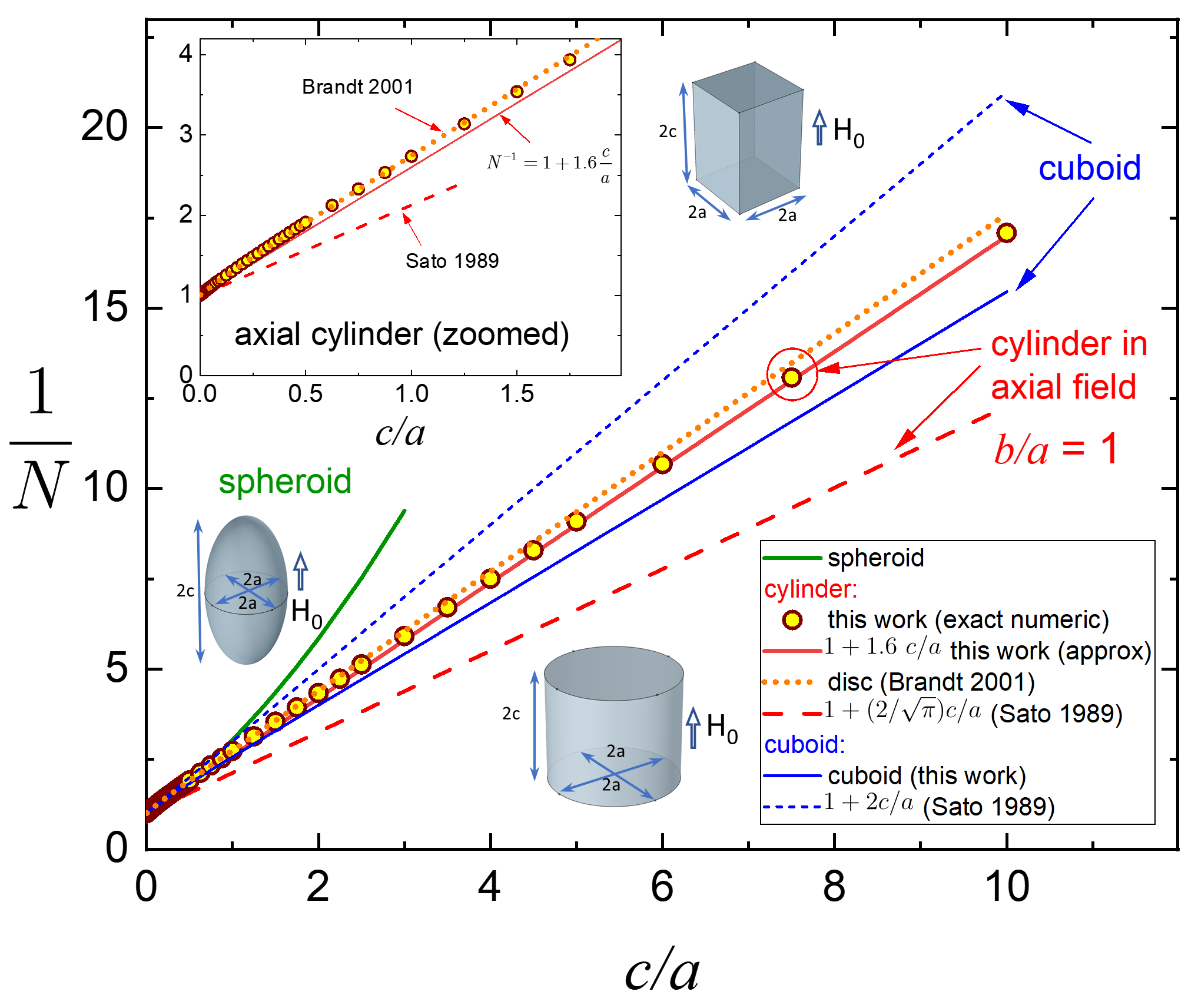}%
\caption{Inverse of the effective demagnetizing factor, $N^{-1}$, as a function of thickness
to diameter aspect ratio for a cylinder. Open symbols are our numerical results and solid line is our approximations, Eq.~(\ref{NaxialCylinder}). For comparison, E.~H.~Brandt's theory is shown by the dotted line and simplified approximations of M.~Sato and Y.~Ischii by the dashed line. For completeness, comparative results for a cuboid are also shown by solid line (this work) and dashed line Ref.~\cite{Sato89}. Inset zooms on to smaller aspect ratio region showing excellent agreement of our numerical results and Brandt's formula \cite{Brandt2001}.
}
\label{fig7}
\end{figure}

This is another typical shape of practical importance and interest. Often it is a piece of a round wire, part of superconducting magnet winding or various cables and transmission lines. They may be subject to either axial or transverse field (or a combination of the two). For the axial case (magnetic field along the cylinder axes) and circular cross-section, the inverse demagnetizing factor is shown in Fig.~\ref{fig7} and compared to square base cuboid and a spheroid of similar aspect ratio. For comparison, Fig.~\ref{fig7} also shows rectangular cuboid and a spheroid as a
function of $c/a$ ratio. Clearly, for a cylinder, Eq.(\ref{NaxialCylinder})
works quite well. Also shown is a comparison with formulas given by Sato and Ishii \cite{Sato89}. They approach our results in the thin limit, but the general trend is quite different.

As shown in Fig.~\ref{fig7}, analysis of the numerical results shows that the simplest curves are obtained for the inverse of the effective demagnetizing factor $N^{-1}$ as a function of the aspect ratio. Indeed, this was also noted in many previous works, for example M.~Sato and Y.~Ishii \cite{Sato89} and E.~H.~Brandt \cite{Brandt2001}. In case of a finite cylinder in axial magnetic field, we obtain a simple approximate formula for the effective inverse demagnetizing factor:
\begin{equation}
N^{-1}_{axial} \approx 1+1.6\frac{c}{a}
\label{NaxialCylinder}%
\end{equation}

We note that our very early work suggested similar approximation with a crude estimate of a numerical factor of 2 in place of 1.6 in Eq.~\ref{NaxialCylinder} \cite{Prozorov2000}. Our results can be compared with the numerical simulations of finite superconducting samples by E.~H.~Brandt \cite{Brandt2001}. He extended his calculations of infinite rectangular strips in perpendicular magnetic field to finite disks of rectangular cross-section. According to Brandt, for a disk of height $2c$ and diameter $2a$ \cite{Brandt2001},

\begin{equation}
N^{-1}_{disk}=1+\frac{1}{q}\frac{c}{a}
\label{Brandt01d}
\end{equation}
\noindent where
\begin{equation}
q=\frac{4}{3\pi}+\frac{2}{3\pi} \tanh{\left[1.27 \frac{c}{a} \ln{\left(1+\frac{a}{c}\right)}\right]}
\label{Brandt01diskq}
\end{equation}

Figure~\ref{fig7} shows excellent agreement between our numerical results and Brandt's equations lending further support to our calculations.
On the other hand, simplified approximations of M.~Sato and Y.~Ishii \cite{Sato89} for these geometries do not agree at all with our and Brandt's results.

\begin{figure}[tb]
\includegraphics[width=1\linewidth]{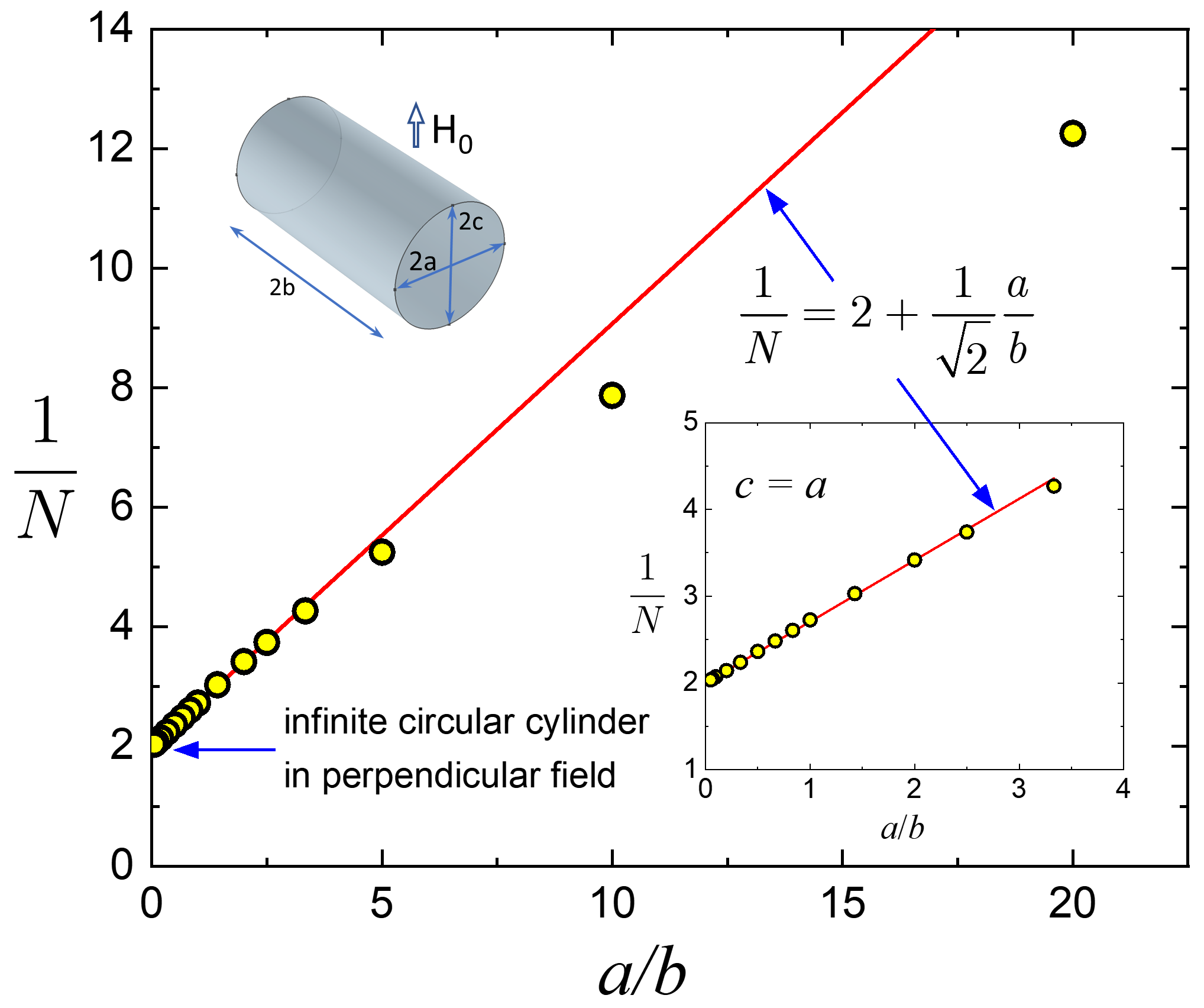}%
\caption{Finite cylinder in transverse magnetic field as a function of ratio of diameter to length. Inset shows small $a/b$ values.}
\label{fig8}
\end{figure}

\subsubsection{Finite circular cylinder in a transverse magnetic field}
We now consider a finite cylinder of circular cross-section ($a=c$) in a magnetic field applied perpendicular to its axis. (Please, note that we change the designations of the dimensions compared to the previous subsection to follow the uniform naming scheme of this paper). Figure \ref{fig8} shows inverse demagnetizing factor $N^{-1}$ as function of the $a/b$ ratio. Notice that this ratio is reciprocal to that of Eq.(\ref{NaxialCylinder}). For small enough ratio of a diameter to length, $a/b$ a good approximation for the demagnetizing factor in this case is:%

\begin{equation}
N^{-1}_{transv} \approx 2+\frac{1}{\sqrt{2}}\frac{a}{b} \label{NtransCyl}%
\end{equation}

Notice that Eq.(\ref{NtransCyl}) gives correct value of $N=1/2$ for an
infinite cylinder in transverse field, $a/b\rightarrow 0$, and correct
$N=0$ when $a/b\rightarrow \inf$.

\subsubsection{Infinite rectangular and elliptical cross-section strips in a transverse field}

Another important case, which is a partial case of the general cuboid is an infinite strip of a rectangular cross-section in a perpendicular field. This geometry is quite relevant for the superconducting tapes as parts of cables or magnet winding. Demagnetizing correction here is an important ingredient of design optimization. It has been considered before by using similar finite element numerical approach as used here, but in two dimensions \cite{Prozorov2000} and also in a different way using highly nonlinear $E(j)$ characteristics applicable for superconductors by H.~Brandt \cite{Brandt2001}.

\begin{figure}[tb]
\includegraphics[width=1\linewidth]{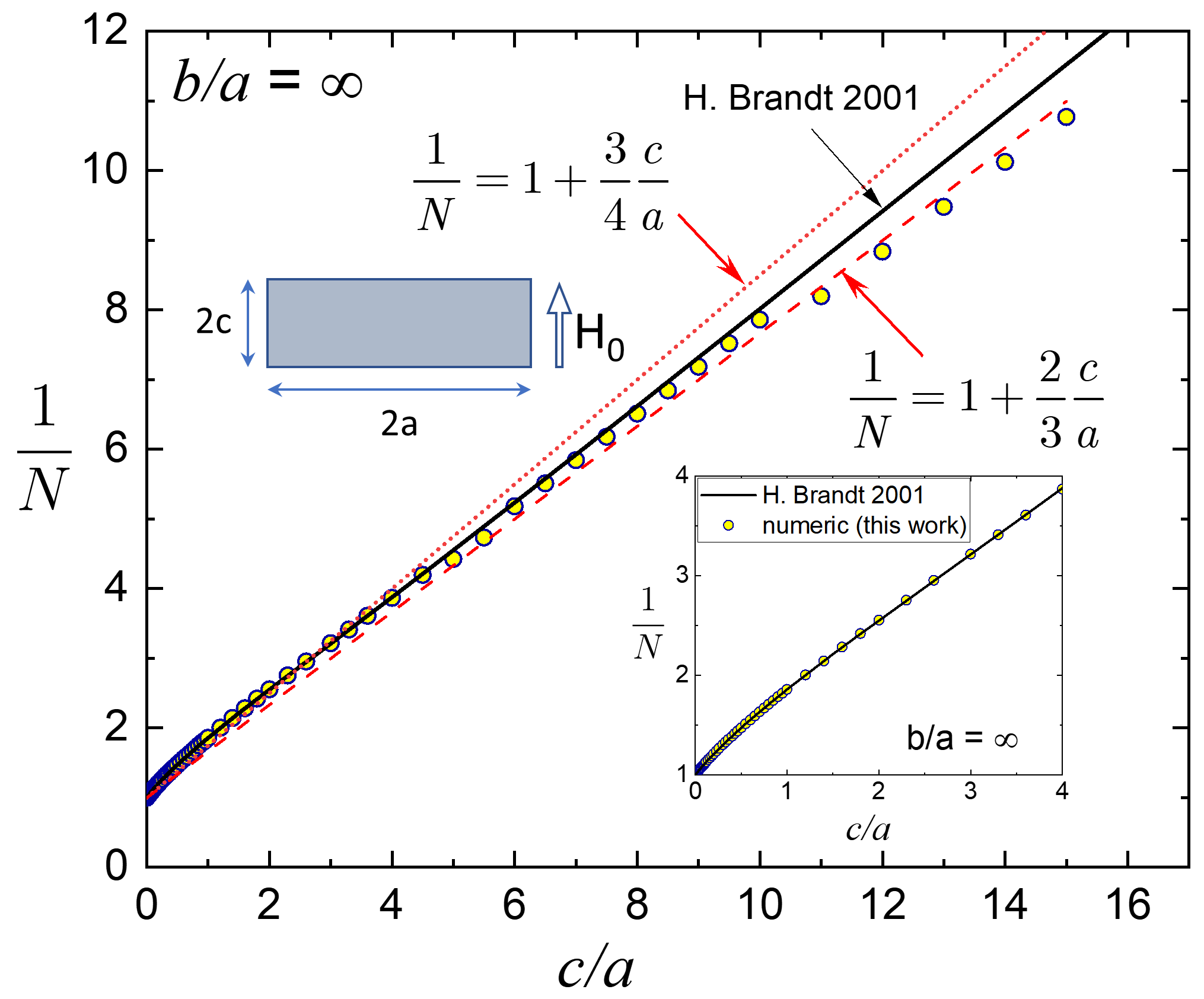}%
\caption{Inverse demagnetizing factor, $N^{-1}$, of an infinite strip of
a rectangular cross-section as function of thickness to width ratio, $c/a$. Two
simple approximate formulas are shown. Also show more elaborate formula by H.~Brandt \cite{Brandt2001}, which agrees with the numerical results quite well up to $c/a=10$. Inset shows smaller range of $c/a$.}
\label{fig9}
\end{figure}

Figure \ref{fig9} shows an inverse demagnetizing factor, $N^{-1}$, of an
infinite strip of a rectangular cross-section as function of thickness to width
ratio, $c/a$. Inset shows smaller range of $c/a$.Two simple approximate formulas are,
\begin{equation}
N^{-1}_{inf-rect-strip}=\left\{
\begin{array}
[c]{c}%
1+\frac{2}{3}\frac{c}{a},\;\text{large }\frac{c}{a} \gtrsim 5\\
\\
1+\frac{3}{4}\frac{c}{a},\;\text{small }\frac{c}{a} \lesssim 5
\end{array}
\right.
\label{NinfStrip}%
\end{equation}

Using numerical simulations of finite superconducting samples, E.~H.~Brandt gives \cite{Brandt2001},

\begin{equation}
N^{-1}_{inf-rect-strip}=1+\frac{1}{q}\frac{c}{a}
\label{Brandt01}
\end{equation}
\noindent where
\begin{equation}
q=\frac{\pi}{4}+0.64 \tanh{\left[0.64 \frac{c}{a} \ln{\left(1.7+1.2 \frac{a}{c}\right)}\right]}
\label{Brandt01q}
\end{equation}

Notably, for a square cross-section ($a=c$) infinite along $b-$direction strip, Brand obtained $N_a=N_c=0.538$, also noting that the sum of demagnetizing factors is greater than 1 for non ellipsoidal shapes (the third, $N_b=0$ for infinite strip).

\begin{figure}[tb]
\includegraphics[width=1\linewidth]{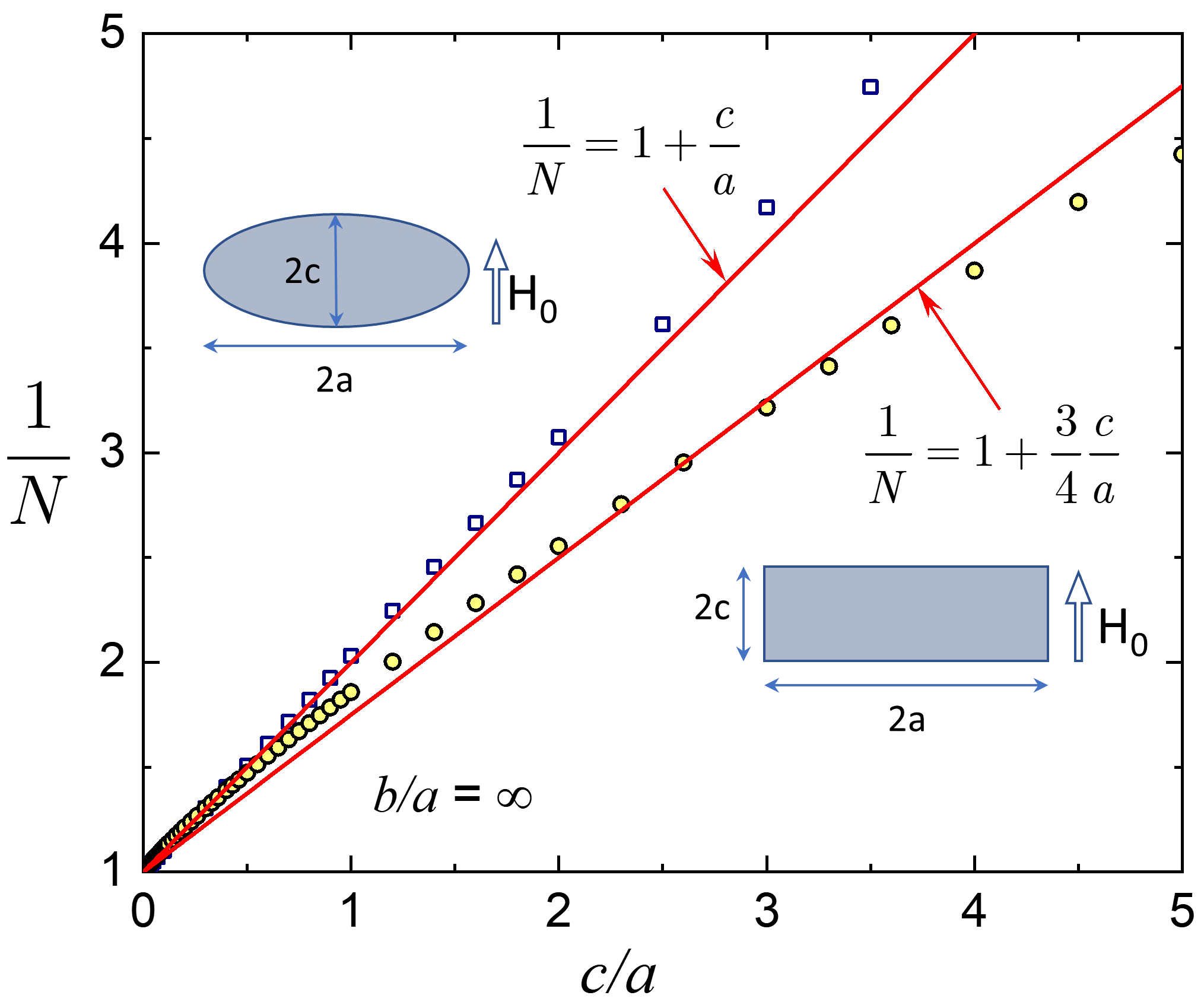}%
\caption{Inverse demagnetizing factor of an infinite strip of elliptical cross-section compared to a
strip of rectangular cross-section. Both are infinite in $b-$direction.}
\label{fig10}
\end{figure}

Next we consider an infinite strip of elliptical cross-section, compared to a rectangular strip in Fig.\ref{fig10}. It turns out, the elliptical cross-section has the simplest approximate equation for the effective demagnetizing factor of all considered cases. Here,
\begin{equation}
N^{-1}_{inf-ell-strip} \approx 1+\frac{c}{a} \label{NinfEllStrip}%
\end{equation}

\subsection{Exotic geometries}

\begin{figure}[tbh]
\includegraphics[width=1\linewidth]{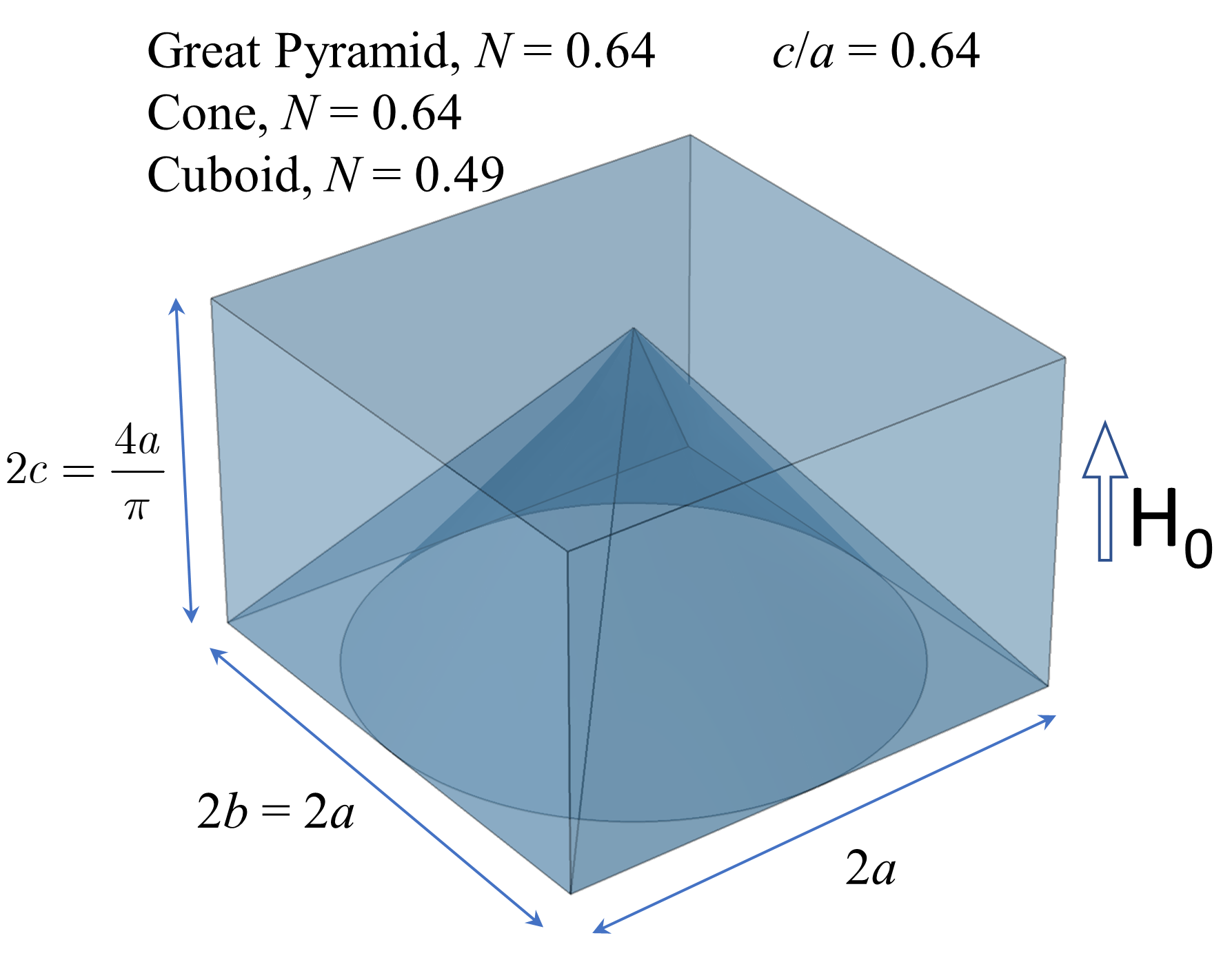}%
\caption{"Exotic" sample shapes: The Great Pyramid of Giza, inscribed cone of the same
height and a cuboid enclosing the pyramid. Effective demagnetizing factor of the pyramid (and the same for the inscribed cone), $N=0.64$, is the same as the ratio of height to side, $c/a=2/\pi=0.64.$. For a cuboid with the same height and side, $N=0.49$.}
\label{fig11}
\end{figure}

Described numerical method allows for the calculation of the effective
demagnetizing correction for samples of any shape. For an illustration, let
us consider a pyramid in shape of the Great Pyramid of Giza, a cone inscribed
in this pyramid and a slab enclosing it, all three shown in Fig.\ref{fig11}.
In all these cases, the ratio of its height to the  side is
$c/a=2/\pi\approx0.64$. Most likely a pure coincidence, but demagnetizing factor of the Great Pyramid (and
of the inscribed cone), $N=0.64$, is the same as the ratio of height to side.
For a cuboid of the same $c$ and $a$, $N=0.49$ is smaller owing this to a larger
volume compared to the cross-sectional area responsible for the magnetic field
distortion around the sample. This adds yet another puzzle for Egyptologists.

\section{Conclusions}

We introduced a direct, assumptions free, efficient way to estimate the effective (or ``integral") demagnetizing factors
of arbitrarily shaped diamagnetic samples. The key equation,  Eq.(\ref{MdefB}), was generalized and re-derived in form of Eq.(\ref{MdefJ}) where vector ${{\cal E}}(\bm \rho)=  \int  d^3\bm r \, {\bm R}/R^3$ can be interpreted as a pseudo-electric field ${\cal E}$ produced by a charge density -1 uniformly distributed in the entire space, and particular form of  ${\cal E}(\bm r)$ depends on the shape of the integration domain. This answers the question posed in the introduction. Namely, it allows calculating total magnetic moment of arbitrarily shaped sample if the volume distribution of the magnetic induction is known. In this work, the latter is obtained from the adaptive finite element full 3D numerical calculations using COMSOL 5.3 software. We provide simple approximate, yet accurate (e.g., for cuboid within $\Delta N < 0.05$, see Fig.~{fig5}(a)) analytical expressions to estimate effective demagnetizing factors of samples of commonly used non-ellipsoidal shapes, which are summarized in Table I.

\begin{acknowledgments}
This work was supported by the U.S. Department of Energy (DOE), Office of Science, Basic Energy
Sciences, Materials Science and Engineering Division. Ames Laboratory is operated for the U.S. DOE by Iowa State University under contract DE-AC02-07CH11358.
\end{acknowledgments}

\bibliographystyle{apsrev4-1}

\clearpage
\cleardoublepage
\appendix
\newpage

\section {Current ring}

To demonstrate how calculations of the total magnetic moment work, it is instructive to consider an example of a ring of a radius $a=1\,$m in plane $z=0$ with current $J/4\pi=1\,$A. The total magnetic moment is $m_z=J \pi a^2 =4\pi^2\,$Am$^2$. The field $B_z$ around the ring according to Landau is \cite{Landau1984}
   \begin{eqnarray}
 B_z=\frac{2}{ \sqrt{(1+r)^2+z^2}}\left[ \bm{K}  +\frac{ 1-r ^2-z^2}{(1-r)^2+z^2}\bm{E} \right],\qquad
  \label{r1}
\end{eqnarray}
\noindent where $r,z$ are cylindrical coordinates, the complete elliptic integrals $\bm{K}, \bm{E}$ are functions of $k^2=4r/[(1+r)^2+z^2]$. The Jackson theorem for this case with no applied field reads:
   \begin{eqnarray}
\frac{2}{3}\,{m}=\int  B_zdV,
  \label{jack}
\end{eqnarray}
 where the integration is over the whole space. Hence, one has to check the equality
   \begin{eqnarray}
\int\limits B_zd^3\bm{r} = \frac{8\pi^2}{3}  \,.
  \label{jack1}
\end{eqnarray}
One can perform the volume integration in spherical coordinates $R,\theta$: $r=R\sin\theta$, $z=R\cos\theta$, $dV=2\pi R^2\sin\theta dR\,d\theta$:
\begin{eqnarray}
  B_z&=&\frac{2}{ \sqrt{1+2R\sin\theta +R^2}} \Big[  \bm{K}  +\frac{ 1-R^2 }{1-2R\sin\theta +R^2}\bm{E} \Big], \nonumber\\
  k^2&=&\frac{4R\sin\theta}{1+2R\sin\theta +R^2}\,.\qquad
  \label{Bz1}
\end{eqnarray}
The integral is readily calculated numerically: $ \int B_z\, d^3\bm{r} \approx 26.32$ coincides with the analytic value $8\pi^2/3$ and gives $m=3I_z/2=4\pi^2\approx 39.48$.\\

Another way of evaluating the integral $\bm I$ in the whole space is to choose cylindrical  integration  elements with axis along the applied field ${\bf H_0}= H_0 {\hat {\bm z}}$. To this end we use Eq.~(\ref{r1}) where $B_z$ is given in cylindrical coordinates. The numerical integration yields $I_z=39.48$ which coincides with the result for spherical integration multiplied by 3/2.   Eq.(\ref{eq9a}) gives  $m_z=I_z=4\pi^2\approx 39.48$. Thus, although the results of numerical evaluation of the integral over the whole space differs from the spherical method, the magnetic moment value for two methods comes out the same.\\

\begin{figure}[tbh]
\includegraphics[width=1\linewidth]{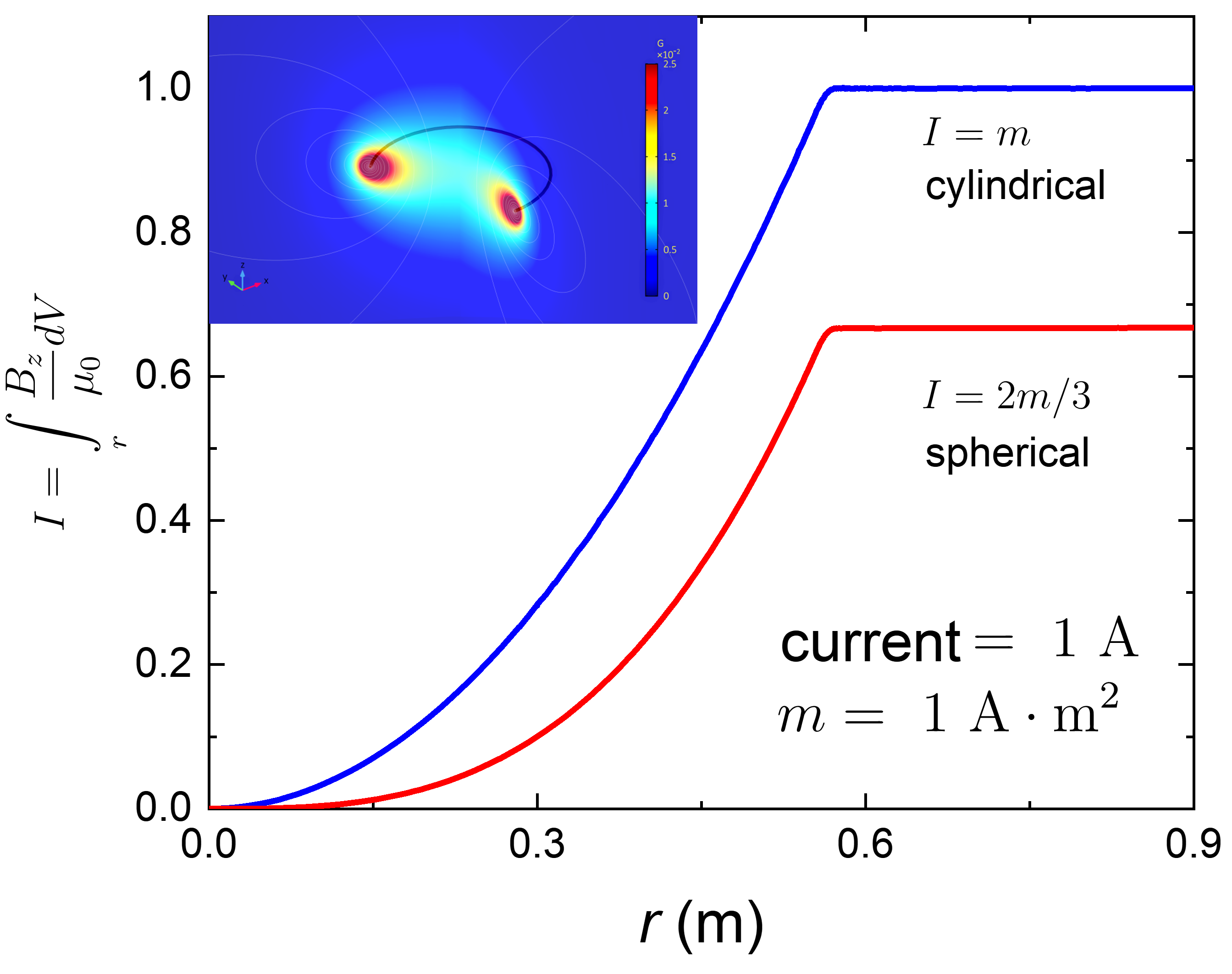}%
\caption{Integral $I_z$, of Eq.~(\ref{eq0}), calculated numerically using spherical shells and cylindrical shells, respectively. The calculations are done in SI units and ring radius ($=1/\sqrt(\pi)$ m) is chosen to give total magnetic moment of 1 Am$^2$ for current of 1 A. This way the integral value is just $1/\alpha$ of Eq.~(\ref{MdefB}). Inset shows three dimensional pie-cut picture of the absolute value of the magnetic induction around the ring.}
\label{fig12}
\end{figure}

To illustrate how our numerical calculations reproduce non-trivial analytical results presented here, Fig.\ref{fig12} shows evaluation of the $z-$component of the integral $I$, Eq.~(\ref{eq0}), using spherical and cylindrical shells. The inset in  Fig.\ref{fig12} shows three dimensional pie-cut picture of the absolute value of the magnetic induction around the ring. The ring radius, $=1/\sqrt{\pi}$ m, is chosen to have ring area to give total magnetic moment of 1 Am$^2$ with 1 A current in the ring. With such choice, the value of the integral $\bm I$ will be equal to $1/\alpha$ of the Eq.~(\ref{MdefB}). Note that numerical calculations here and everywhere in this paper are carried out in SI units. We obtain that the spherical shell integration tends to the value of $\alpha^{-1}=2/3$ and cylindrical shell integration to $\alpha^{-1}=1$, exactly as shown analytically. Moreover, both integrals stop changing as soon as the sample current is fully enclosed in the integration volume, again as expected from the analytical calculations. Therefore, integration of space outside the sample does not contribute to the integral $\bm{I}$. It does not mean, however, that the outer space can by truncated for numerical calculations. It should still be much larger compared to the sample size in order to solve for (very long - range) magnetic fields distribution correctly.

 \section{Identity Eq.~(\ref{MdefB}) in Gaussian units}

  The derivation of this identity is given in the main text in the ``recommended" SI system, which was also used in COMSOL to solve the Maxwell equations for the field distribution in SI.  This, however, makes the derivation unnecessary cumbersome. Many researchers working in the fields of superconductivity and magnetism prefer CGS and it is also used in magnetometers such as \emph{Quantum Design} MPMS. We therefore provide here the same derivation in Gaussian system, which, in our opinion, makes the derivation more transparent. \\

The total magnetic moment $\bm m$ of a finite-size sample is proportional to the volume integral of the total field distribution $\bm B(\bm r)$ over the sphere of large enough radius ${\cal R}_1$ such that the whole sample is situated inside the sphere. This statement is proven
 in Jackson's book \cite{Jackson2007} which gives:
  \begin{eqnarray}
 { \bm I }=\int_{r<{\cal R} }\left[{\bm B}(\bm r)-{\bm H}_0\right]d^3\bm r =\frac{8\pi}{3}\,\bm m\,.
  \label{eq0}
\end{eqnarray}
Here, ${\bm H}_0$ is the   applied   field. In particular, one can take ${\cal R} \to\infty$, i.e. the integral can cover the whole space.
This identity is proven in Ref. \,\onlinecite{Jackson2007} by expanding the field outside a sphere $\cal R$ which contains the sample (the currents)  in spherical harmonics. From the point of view of a finite-element numerical method it is is more convenient to deal with cylindrical integration domains. We thus provide a proof   not related to a particular shape of the integration region.

The field $\bm B$ consists of the applied field and the field $\bm b$ due to currents $\bm J$ in the sample of a finite volume $V$:
 \begin{eqnarray}
{\bm B}={\bm H}_0 +\bm b\,,
  \label{eq1}
\end{eqnarray}
i.e.,  $\bm I=\int {\bm b}\, d^3\bm r$, where according to Biot-Savart
 \begin{eqnarray}
  {\bm b}(\bm r)=\frac{1}{c}\int_V d^3{\bm\rho}\,\frac{{\bm J}(\bm \rho)\times{\bm R}}{R^3}\,,\qquad \bm R=\bm r -\bm \rho\,.\qquad
  \label{eq2}
\end{eqnarray}
Hence, we have
 \begin{eqnarray}
 c\bm I&=& \int   d^3\bm r \int_V d^3{\bm\rho}\frac{{\bm J}(\bm \rho)\times{\bm R}}{R^3}\nonumber\\
 &=&
\int_V d^3{\bm\rho} \,{\bm J}(\bm \rho)\times \int  d^3\bm r \frac{{\bm R}}{R^3} \,,
  \label{eq3}
\end{eqnarray}
where the integration over $\bm r$ is extended to the whole space.\\

  The vector ${\bf{ \cal E}}(\bm \rho)=  \int  d^3\bm r  {\bm R}/R^3$ is analogous to the electrostatic field of a charge distribution with   density of $-1$ in the whole space.
 For such a distribution, the   field  ${ \cal E}$ is not defined uniquely, it depends on the way one divides the space in charged elements. \\

 For $\bm\rho=0$, we must have  ${\bf { \cal E}}=\int d^3\bm r  ({\bm r}/r^3)=0$  by symmetry.
If one uses elements as spherical shells, and applies the Gauss theorem to a sphere of a radius $\rho$ one obtains:
 \begin{eqnarray}
{ \cal E}=-\frac{4\pi}{3}\,\bm\rho \,.
  \label{eq8}
\end{eqnarray}
Hence, we have
 \begin{eqnarray}
  \bm I= -\frac{ 4\pi }{3c}\int_V d^3{\bm\rho} \,{\bm J}(\bm \rho)\times {\bm \rho}=\frac{8\pi}{3} {\bm m} \,,
  \label{eq9}
\end{eqnarray}
where $\bm m$ is the total  magnetic moment. It is worth noting that this formula holds for any current distribution within the finite sample of arbitrary shape. \\

If one uses elements as cylindrical shells parallel to $\bm H_0$, i.e. choose the volume element as $2\pi \rho_1\,d\rho_1dz$ ($\bm \rho_1$ is the cylindrical radius vector), and applies the Gauss theorem to a cylinder of a radius $\rho$ one obtains:
 \begin{eqnarray}
{ \cal E}  =- 2\pi {\bm \rho}_1 \,.
  \label{eq8a}
\end{eqnarray}
where $\bm\rho_1$ is now the cylindrical radius vector.
Substituting this in Eq.\,(\ref{eq3}), one expresses the $z$ component of the integral $\bm I$:
 \begin{eqnarray}
   I_z=  4\pi  \,   m_z \,.
  \label{eq9a}
\end{eqnarray}

\clearpage
\cleardoublepage
\appendix
\newpage
\onecolumngrid
\appendix{Appendix: Summary of approximate formulas for the effective demagnetizing factor $N$}%

\begin{table}
\caption{Approximate effective (``integral") demagnetizing factor along applied magnetic field.}
\label{TableDemag}

\begin{tabular}{m{4cm}m{4cm}m{10cm}}
\hline
\large{Shape} & \large{Geometry} & \large{Demagnetizing factor along applied field}\\
\hline
\hline
Ellipsoid (exact) &\includegraphics[width=2.2cm]{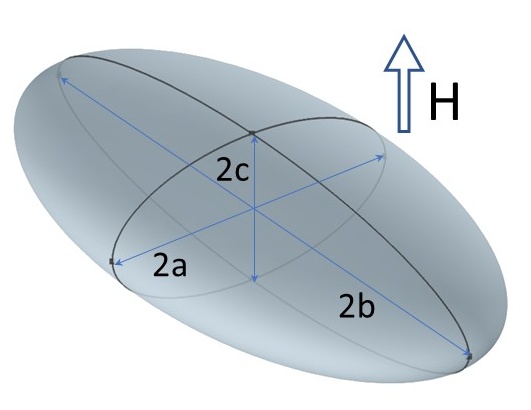} & $\large{N}=\frac{1}{2}\frac{b}{a}\frac{c}{a}\displaystyle\int\limits_{0}^{\infty}\frac{ds}{(s+\frac{c^2}{a^2})\sqrt{\left(  s+1\right)  \left(  s+\frac{b^2}{a^2}\right)  \left(  s+\frac{c^2}{a^2}\right)}}$  \\
Rectangular cuboid &\includegraphics[width=2.2cm]{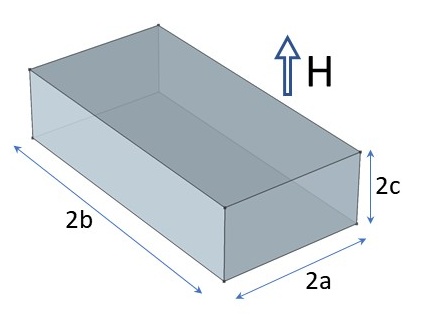} & \large{$N^{-1}=1+\frac{3}{4}\frac{c}{a}\left(1+\frac{a}{b}\right) $}  \\
Strip, rectangular &\includegraphics[width=2.2cm]{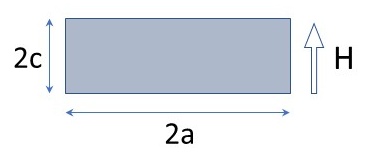} & \large{$N^{-1}=1+\frac{2}{3}\frac{c}{a}$} \small{for $(c/a \gtrsim 5)$},
\large{$=1+\frac{3}{4}\frac{c}{a}$} \small{for $(c/a \lesssim 5)$}  \\
Strip, elliptical &\includegraphics[width=2.2cm]{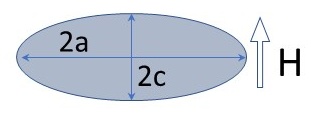} & \large{$N^{-1}=1+\frac{c}{a}$}  \\
Cylinder, axial &\includegraphics[width=2.2cm]{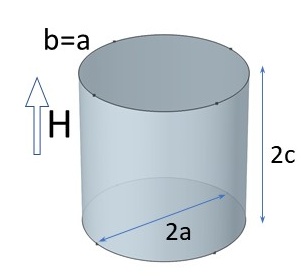} &   \large{$N^{-1}=1+1.6\frac{c}{a}$}\\
Cylinder, transverse &\includegraphics[width=2.2cm]{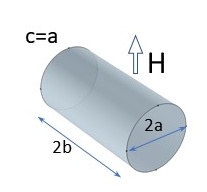} &  \large{$N^{-1}=2+\frac{1}{\sqrt{2}}\frac{a}{b}$} \\
\hline
Great pyramid\\ (cone, cuboid) &\includegraphics[width=2.7cm]{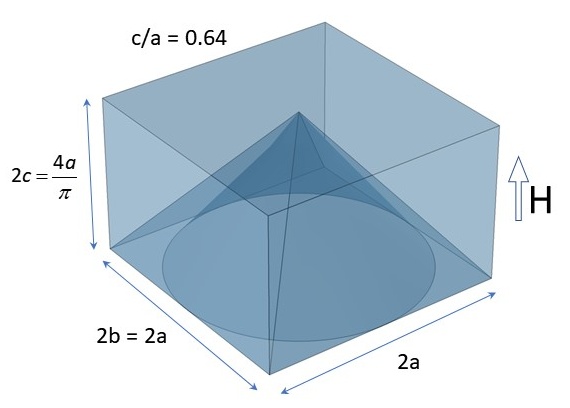} & $N=0.64$ $(0.64, 0.49)$  \\
\hline
\end{tabular}
\end{table}

\end{document}